\documentclass[a4paper,fleqn,usenatbib]{mnras}
\usepackage[T1]{fontenc}
\usepackage{ae,aecompl}

\usepackage{graphicx}	% Including figure files
\usepackage{amsmath}	% Advanced maths commands
\usepackage{amssymb}	% Extra maths symbols

\newcommand{\Msun}{\ifmmode {M_{\odot}}\else${M_{\odot}}$\fi}
\newcommand{\Lsun}{\ifmmode {L_{\odot}}\else${L_{\odot}}$\fi}
\newcommand{\Rsun}{\ifmmode {R_{\odot}}\else${R_{\odot}}$\fi}
\newcommand{\dt}{\ifmmode {\Delta \text{Log}T_{\text{eff}}}\else{$\Delta$Log$T_{\text{eff}}$}\fi}
\newcommand{\lt}{\ifmmode {\text{Log}T_{\text{eff}}}\else{Log$T_{\text{eff}}$}\fi}

\title[soft spectrum of V0332+53]{The Soft X-ray Spectrum of the High Mass X-Ray Binary V0332+53 in Quiescence}

\author[Elshamouty et al.]{
Khaled G.~Elshamouty,$^{1}$\thanks{E-mail: alshamou@ualberta.ca}
Craig O.~ Heinke$^{1}$,
and Rhys Chouinard$^{1,2}$
\\
$^{1}$Physics Department, University of Alberta, AB, Canada\\
$^{2}$City of St. Albert, AB, Canada\\
}

\date{Accepted XXX. Received YYY; in original form ZZZ}

\pubyear{2016}

\begin{document}
\label{firstpage}
\pagerange{\pageref{firstpage}--\pageref{lastpage}}
\maketitle

\begin{abstract}
The propeller effect should cut off accretion in fast-spinning neutron star high mass X-ray binaries (HMXBs) at low mass transfer rates. However, accretion continues in some HMXBs at $L_{\textsl{x}} < 10^{34}$ erg s$^{-1}$, as evidenced by continuing pulsations. Indications of spectral softening in systems in the propeller regime suggest that some HMXBs are undergoing fundamental changes in their accretion regime. 
A 39 ks \textit{XMM-Newton} observation of the transient HMXB V0332+53 found it  at a very low X-ray luminosity ($L_{\textsl{x}} \sim 4\times 10^{32}$ erg s${^{-1}}$).  A power-law spectral fit requires an unusually soft spectral index ($4.4^{+0.9}_{-0.6}$), while a magnetized neutron star atmosphere model, with temperature \lt\  6.7$\pm 0.2$ K and inferred emitting radius of $\sim0.2-0.3$ km, gives a good fit. We suggest that the quiescent X-ray emission from V0332+53 is mainly from a hot spot on the surface of the neutron star.  We could not detect pulsations from V0332+53, due to the low count rate.  Due to the high $N_H$, thermal emission  from the rest of the neutron star could be only weakly constrained, to \lt\ $<$6.14$^{+0.05}_{-6.14}$ K, or $<3\times10^{33}$ erg s${^{-1}}$. 
\end{abstract}


%%%%%%%%%%%%%%%%% INTRODUCTION %%%%%%%%%%%%%%%%%%

\section{Introduction}

Accreting X-ray pulsars (XRPs) in high mass X-ray binaries (HMXBs) provide an excellent laboratory for accretion physics. The strong ($B\sim$10$^{11-13}$ G) neutron star (NS) magnetic field means that the NS's magnetosphere determines the conditions of the accretion flow. The magnetospheric radius is defined to be where the pressure of the magnetic field $B(r_{m})^{2}/8\pi$ equals the ram pressure of the infalling matter, $\rho(r_{m})v(r_{m})^{2}$. At this radius, matter is thought to thread itself onto magnetic field lines. If this magnetospheric radius becomes larger than the corotation radius $R_{c} = (GMP^{2}/4\pi^{2})^{1/3}$, where the Keplarian angular velocity of infalling matter matches the angular velocity of the spinning NS, the infalling matter will not be moving fast enough to join the magnetic field lines \citep{Illarionov75}. This is the \textit{propeller} regime, where the accretion is stopped, or at the very least greatly reduced, when the accretion rate falls below the minimum threshold value of 
\begin{equation}
L_{\textsl{min}} = 1.6 \times 10^{37} R_{12}^{2} M_{1.4}^{-2/3}B_{12}^{2}P_{1}^{-7/3} \textsl{erg s}^{-1}
\end{equation}\label{eq:lmin}
where $B_{12}$,$R_{12}$,$M_{1.4}$ and $P_{1}$ are the NS's surface magnetic field in the units of $10^{12} G$, radius in units of 12 km, mass in units of 1.4 $M_{\odot}$ and spin period in units of seconds, respectively \citep{Stella86}. \\

Rapid drops in $L_{\textsl{x}}$ by a factor $>100$ during the late stages of an outburst have been seen in several systems, and interpreted as marking the transition to the propeller regime. Some examples include XRPs in HMXBs like 4U 0115+63 and V0332+53 \citep{Campana01,Stella86,Tsygankov16},  low magnetic field ($B\sim10^8-10^9$ G) fast-spinning NSs in low-mass X-ray binaries such as Aquila X-1  \citep{Campana98b,Campana14}, SAX J1808.4-3658 \citep{Campana08}, and XTE J1701-462 \citep{Fridriksson10}, and the  bursting pulsar GRO J1744-28, with an intermediate-strength ($B\sim5\times10^{11}$ G) magnetic field \citep{Cui97}. 
Understanding the transition to the propeller regime is important for understanding the luminosity function of X-ray binary populations \citep{Shtykovskiy05} and the spin-up of millisecond radio pulsars (e.g. \citealt{Tauris12}, \citealt{Papitto14}, \citealt{Archibald15}).

For a system in the propeller regime, there are a few candidates for the mechanism of continued X-ray emission. If the propeller effect is not 100\% effective, calculations have suggested that a small amount of matter ($\sim 1\%$) may penetrate the magnetosphere, and reach the surface of the NS producing  X-ray emission, while the majority of matter is ejected (e.g. \citealt{Stella86}). Magnetohydrodynamics simulations (e.g. \citealt{Romanova04}) agree with this basic picture, providing a possible explanation for continued luminosity in quiescence, which is predicted to be pulsed (see below). 

This low-level accretion luminosity may show a similar spectrum to that at higher $L_{\textsl{x}}$, or resemble a hot, small-radius blackbody, indicating hot polar caps, as seen in some radio pulsars (e.g. \citep{Zavlin98,Ozel13}). Emission from the hot neutron star surface will generate a blackbody-like thermal spectrum, shifted slightly to higher energies by passage through the hydrogen atmosphere, depending on the magnetic field strength \citep{Pavlov95}. The neutron star core is expected to be heated by accretion during outbursts, providing a steady-state heat flux to the surface during quiescence \citep{Brown98}. Alternatively, X-ray flux might be generated by accretion down to the magnetosphere, providing a luminosity a factor $L(r_{m})/L(R) = 2r_{\textsl{cor}}/R_{\textsl{NS}} \sim 1000$ lower than accretion onto the neutron star surface, and unpulsed (e.g. \citealt{Campana98a}).  However, accretion down to the magnetosphere is not certain to produce X-rays, and can be ruled out for some XRPs in quiescence (e.g. A 0535+26, \citealt{Negueruela00}).  Finally, in some cases the O or B companion star may generate soft X-rays at the level of 10$^{31-33}$ erg s$^{-1}$, which may be detectable in deep quiescence as a line-dominated thermal spectrum. Accurate measurements of the spectrum and pulse fraction can determine what role each of these play in the quiescent emission.

The heating of the NS core during outbursts should lead to thermal X-ray emission from the entire NS surface in quiescence, unless strong neutrino emission is present \citep[e.g.][]{Yakovlev03,Heinke07,Wijnands13}.  
The luminosity expected depends on the long-term average accretion history, and on the mass of the neutron star, and may lie in the range $\sim$10$^{31}$--10$^{33}$ erg s$^{-1}$. 

The magnetic fields are known (by X-ray cyclotron line measurements) for some XRPs in HMXBs, and some XRPs have been observed at $L_{\textsl{x}}$ that are well within the propeller regime. 
Three HMXBs with known B fields, distances and spin periods have X-ray  measurements that are clearly in the propeller regime: A0535+262 \citep{Negueruela00,Mukherjee05,Rothschild13,Doroshenko14}; V0332+53 \citep{Stella86,Campana02}; and 4U 0115+63 \citep{Campana01,Campana02}.  Observations of A0535+262 with various telescopes showed clear pulsations and a spectrum describable by a cutoff power-law (photon index $\sim$0.4, $E_{fold}\sim4$-6 keV)  while in the propeller regime \citep[e.g.][]{Doroshenko14}.
The other two systems, which are significantly more distant, have only weak constraints on their quiescent spectra and pulsations from BeppoSAX \citep{Campana02}.
Several other systems have published observations which have been claimed to likely be in the propeller regime, but in these cases the magnetic field is less certain (e.g. A 0538-668, \citealt{Campana02}; GX 1+4, \citealt{CuiSmith04}; Vela X-1, \citealt{Doroshenko11}). The systems 4U 1145-619 \citep{Mereghetti87,Rutledge07} and 1A 1118-615 \citep{Rutledge07} have been observed to show pulsations and hard power-law spectra when $L_X\sim2-3\times10^{34}$ erg/s, with photon indices $\sim$0.8--1.6.  

Several other XRP systems show evidence of thermal blackbody-like emission while at low X-ray luminosities. SAX J2103.5+4545 \citep{Reig14} demonstrated X-ray pulsations and a $\sim$1 keV blackbody thermal spectrum at $L_X=1.2\times10^{33}$ erg s$^{-1}$.  No variation other than the pulsations were seen. It is unclear if this NS is in the propeller state, since the B field is not accurately known, but it was suggested that the observed X-rays are from hot spots still cooling from the latest outburst. Soft thermal components to XRP X-ray spectra around $10^{34-35}$ erg s$^{-1}$ have been observed from  RX J0146.9+6121 \citep{LaPalombara06}, 4U 0352+309/ X Per \citep{LaPalombara07}, and RX J1037.5-5647 \citep{LaPalombara09}, suggesting the presence of hot spots on NSs. In no low-luminosity XRP X-ray spectrum has a thermal component consistent with emission from the full NS surface been unambiguously detected.

V0332+53 is a transient XRP in a 33.85-day orbit \citep{Doroshenko16} with a Be-type companion, spectroscopically classified as O8-9Ve \citep{Negueruela99}. The NS spin period is 4.35 s \citep{White83}, and its magnetic field has been measured (through detection of a cyclotron line) to be $2.7\times10^{12}$ G \citep{Makishima90,Pottschmidt05}. 
It is thought to be at a distance of roughly 7 kpc (a plausible range includes 6-9 kpc), based on spectral typing of the companion star \citep{Negueruela99}.
V0332+53 was first detected during a large outburst in 1973 by the \textit{Vela 5B} satellite \citep{Terrell84}. It has since been observed in outburst in 1983 \citep{Tanaka83,Stella85}, in 1989 \citep{Makishima90}, in 2004 \citep{Kreykenbohm05,Pottschmidt05}, in late 2008 \citep{Caballero-Garcia15}, and in 2015 \citep{Tsygankov16,Wijnands16}. 
Between outbursts, V0332+53 still gives off X-rays, but at only a small fraction of its outburst luminosity \citep{Campana02}.  We proposed an XMM-Newton observation to determine the shape of its spectrum in quiescence, and search for pulsations.  

\section{Spectral Analysis}

We observed V0332+53 with \textit{XMM-Newton} for 39 ks on Feb. 10th, 2008 (ObsID 0506190101).  This is at orbital phase 0.7 after periastron, according to the ephemeris of \citet{Doroshenko16}. Here we discuss only the data obtained with the pn \citep{Struder01} and MOS \citep{Turner01} cameras. We used SAS v13.5.0 software package to reduce the data. We filtered the data to the range of 0.2 -- 10 keV for all three cameras. We created lightcurves from all data from each camera, binned to 50 second bins, to  filter out times of high background rates due to soft proton radiation. We removed times with event rates higher than 0.16, 0.12 and 0.2 counts/s for the MOS1, MOS2 and pn cameras respectively, which removed 2 ks from the observation time. We used a circle of radius $10\arcsec$ to extract the source, and a  $15\arcsec$ circle for the background. We created appropriate response files and effective areas for each camera, for source and background.

We  grouped the data at three different trial values of 15, 20 and 25 counts/bin. We adopted a grouping of 20 counts/bin for our analysis for the three cameras. We combined the extracted spectra from MOS1 and MOS2 into one spectrum and fit it simultaneously with the spectrum extracted from the pn camera, using several different models (\texttt{nsa, nsa[pole], nsa[pole]+pow, pow, nsa[pole]+nsa}), with the details of the spectral fits listed in Table 1, and discussed below. All models included extinction, using the \texttt{tbabs} model, with {\it wilm} abundances \citep{Wilms00}, which we left a free parameter.
The \texttt{nsa} model is a neutron star atmosphere model, including appropriate physics for magnetic fields of $10^{12}$ or $10^{13}$ G \citep{Pavlov95}. We fixed the magnetic field  at $B = 10^{12}$ Gauss for our fits.
%{\bf Check effects if use $10^{13}$, as we are in between these two B values.} 
In all fits using \texttt{nsa}, we fixed the NS mass to $1.4$ \Msun, and the NS radius to $11$ km. We froze the normalization value ('norm', which gives the angle on the sky) to $K = 2.04 \times 10^{-8}$ pc$^{-2}$ (implying an 11 km NS and a distance of 7.0 kpc) when we were modeling emission from the entire star with the \texttt{nsa} model. When we modeled emission from polar caps (we refer to this as \texttt{nsa[pole]} models), we allowed this parameter to vary.

We may predict $N_{H}$ from the optical extinction, using $N_{H} = (E_{B-V})(R_V)(2.8\times10^{21})$ \citep{Bahramian15,Foight15}. For V0332+53, $E_{B-V}$ has been measured by the strength of diffuse interstellar bands to be 1.88$\pm0.1$ \citep{Corbet86}, and, by comparing the spectral classification with the observed colour, to be 1.87 \citep{Negueruela99}. We can then estimate $N_{H} = 1.6 \times10^{22}$ cm$^{-2}$. However, X-ray observations have given values of $N_H$ varying between $6\times10^{21}$ and $1.5\times10^{22}$ cm$^{-2}$ \citep[][note these $N_H$ measurements were on a different abundance scale]{Stella85}, or between $1.0-1.3\times10^{22}$ cm$^{-2}$ \citep{Tsygankov16}. Possibly the extinction may be smaller towards the NS vs. its companion (not implausible, given the excretion disk). We left $N_H$ a free parameter.

\begin{table*}
	\centering
	\begin{tabular}{lllllll}
		\hline
		Parameter/Model                       		   	& pow             			& nsa        	& nsa[pole]           		& nsa[pole]+pow 		&  \multicolumn{2}{|c|}{nsa+nsa[pole]} \\
		\hline
		$N_{H} [\times 10^{22} \text{cm}^{-2}]$ 	   	
		&  $2.6^{+0.8}_{-0.6}$   	&  $3.34$   	&   $1.00^{+0.45}_{-0.31}$&   $1.19^{+0.58}_{-0.45}$&   2.06$^{+0.62}_{-1.19}$& - \\
		Log$T_{\textsl{eff}}[K]$                 	 	   	& ...             			& $6.25$    	& $6.77^{+0.07}_{-0.10}$ 	& $6.69^{+0.13}_{-0.17}$  & 6.14$^{+0.05}_{-6.14}$  & 6.82$^{+0.11}_{-0.12}$  \\
		 $\chi^{2}_{\nu}/$dof                  			   	& 0.81/20        			& $2.60/20$  	& $0.87/19$           		& $0.88/18$             		& 0.80/18 				& - \\
		$\Gamma$                             				&  4.4$^{+0.9}_{-0.7}$   	& ...    		& ...                  			& (1.5)               		& ... 					& ... \\
                $R_{emit}$, km                        				& ...             			&  (11)    		& $0.27^{+0.22}_{-0.10}$  &  $0.36^{+0.72}_{-0.17}$     & (11)  & 0.21$^{+0.21}_{-0.12}$  \\
	$L_{\textsl{x,unabs}} [\text{erg s}^{-1}]$ & $4.9^{+1.0}_{-1.0} \times 10^{33}$ & $3.5\times 10^{33}$ & $3.7^{+1.1}_{-0.5}\times 10^{32}$ & $4.4^{+2.1}_{-0.6}\times 10^{32}$  &  $1.07^{+0.94}_{-1.07}\times 10^{33}$ & $3.6^{+1.2}_{-1.1}\times 10^{32}$  \\
        $L_{bol,th} [\text{erg s}^{-1}]$            				& ...             & $5.2\times 10^{33}$ & $3.9^{+1.3}_{-0.6}\times 10^{32}$ & $4.0^{+2.8}_{-0.6}\times 10^{32}$ & $1.96^{+1.33}_{-1.96}\times 10^{33}$ & $3.7^{+1.4}_{-1.2}\times 10^{32}$  \\
		\hline
	\end{tabular}\label{models}
\caption{Spectral analyses of V0332+53. Errors are 90\% confidence, and are not calculated if $\chi^2_{\nu}$ exceeds 2.0. $L_X$ is given for the 0.5-10 keV range, while $L_{bol,th}$ gives the 0.1-10 keV luminosity for thermal components, both for an assumed distance of 7 kpc. $R_{emit}$ is a rough estimate of the emitting radius, defined as the square root of the {\texttt nsa} normalization times distance (in parsecs), multiplied by the assumed NS radius.  }
\end{table*}

A power-law model, typical of HMXB outburst spectra, provides a reasonable fit, with a reduced chi-squared of 0.81, but an extremely steep best-fit photon index of 4.4$^{+0.9}_{-0.7}$, strongly implying that the emission is thermal in origin. 
The \texttt{nsa} model also does not produce an adequate fit, with a reduced chi-squared of 2.6 for 20 degrees of freedom.
  This suggests that we do not see emission from the neutron star's entire surface. We then consider emission from a small region on the surface of the neutron star, the heated polar cap(s), by allowing the \texttt{nsa} normalization to be free (the \texttt{nsa[pole]} model). This gives a reasonable fit, with a reduced chi-squared of 0.98. The $N_{H}$ in this model is constrained to 0.7$\pm 0.2 \times 10^{22}$ cm$^{-2}$, which is interestingly low compared to the optical estimate. 
The \texttt{nsa[pole]} model gives a rough estimate of the radius of the emitting region of $R_{\textsl{spot}}$ = 0.27$^{+0.22}_{-0.10}$ km, which may suggest the radius of the heated polar cap. This is comparable to, if a little larger than, the predicted radius for the polar cap of a pulsar with spin period of 4.35 seconds and radius 11 km, of 80 m \citep{Lyne06}.  Note that even if the poles were illuminated only over an 80-m region, a larger portion of the surface may be heated by conduction, as suggested for several pulsars \citep[e.g.][]{Zavlin98,Zavlin06}, though the high magnetic field will limit the thermal conductivity perpendicular to the B field lines \citep{Hernquist85}.

We also test two models adding additional, physically motivated components to the {\texttt nsa[pole]} model.
We first tried adding a power-law component (with fixed $\Gamma = 1.5$, due to our limited statistics) to the \texttt{nsa[pole]} model.  Such a power law might be predicted for  X-ray emission from continuing accretion, as seen in low-mass X-ray binaries, and suggested as a possible spectral shape during the late decay of V0332+53's 2015 outburst \citep{Wijnands16}. This also gives an adequate fit, though not a better one (one bin is marginally improved), and an F-test indicates that adding the additional component is not justified for the improvement in spectral quality. 
%This fit permits a larger inferred spot radius, $R_{\textsl{spot,pow}} = 490$ m, 
We also added thermal emission from the entire NS, thus a {\texttt nsa[pole]+nsa} model. This fit, although more of an improvement, was also not justified in terms of the quality of the spectral fit (an F-test gives an F-statistic of 2.79, and probability 0.11 of obtaining such a spectral improvement by chance), but it is useful for permitting a constraint upon thermal emission from the NS (see the Discussion). The spectra, fit to this model, are shown in Fig.~\ref{fig:spec}. 

For completeness, we also tested how our results would be affected if we used the $B=10^{13}$ G {\texttt nsa} model, instead of the $10^{12}$ G model. The {\texttt nsa[pole]}, $10^{13}$ G model still provides a good fit ($\chi^2_{\nu}$=0.82), with a smaller temperature (log $T_{eff}$=6.56$^{0.53}_{-0.38}$) and a larger radius (0.9 km).

\begin{figure}
	\includegraphics[width=0.5\textwidth]{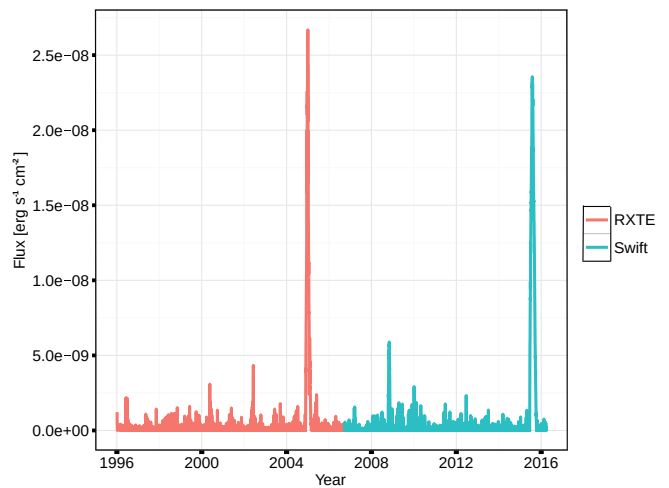}
    \caption{RXTE/ASM and Swift/BAT lightcurves for V0332+53. Daily measurements for both were converted to Crab units, and assuming a power-law spectrum between 0.1 keV - 100 keV, with photon index of 2 converted to bolometric fluxes. }
    \label{fig:lc}
\end{figure}

\begin{figure}
	\includegraphics[width=0.45\textwidth]{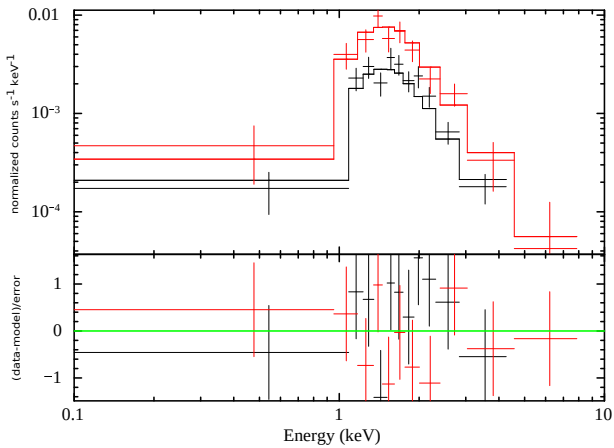}
    \caption{Fit of all spectra from MOS1+MOS2 (combined; black) and PN (red) to the {\texttt nsa[pole]+nsa} model, representing emission from a hot spot, and from the full surface, of a NS, with $B=10^{12}$ G.}
    \label{fig:spec}
\end{figure}

\section{Pulsation Search} 

We attempted to search for pulsations from V0332+53, which is feasible for XMM imaging observations given the relatively long known spin period of 4.35 seconds. We extracted lightcurves from 15\arcsec\ radius circular regions in the MOS and pn data, within which 239 photons were found, of which we estimate 55 to be background. We searched for periodicities using the {\it powspec} command in FTOOLS.  No significant periodicities were detected.  

To estimate our sensitivity limit for pulsations, we simulated XMM-Newton observations with different pulse fractions.  We defined the pulse fraction as the semiamplitude of modulation, divided by the average source count rate \citep{Campana01}.  We tried various pulsed fractions up to 95\%, but were unable to generate a detectable signal in these simulations.  Trials with 1000 simulated source photons (vs. the 184 we obtained) produced a signal when the pulsed fraction reached 45\%. We conclude that our observations were completely insensitive to pulsations from V0332+53 in quiescence, and that such searches on low-luminosity targets are quite difficult.

\section{Discussion}

For V0332+53, the value of the propeller regime cutoff luminosity from eq.~\ref{eq:lmin} is $L_{\textsl{min}}\sim 3\times 10^{36}$ erg s$^{-1}$, for typical NS mass and radius estimates. In table~\ref{models}, we find $L_X$ estimates, for different model choices, to be 0.4-1.4$\times10^{33}$ erg s$^{-1}$. This is one of the lowest luminosities yet measured for a HMXB in quiescence, and is well below the cutoff luminosity for the propeller regime. \citet{Tsygankov16} argue that V0332+53 showed clear evidence of a propeller effect truncating accretion during its 2015 outburst, while \citet{Wijnands16}, observing the late decay with Swift, identify  a slow dimming trend, suggesting either low-luminosity continuing accretion, or cooling of the NS crust post-accretion.  Our observation of purely thermal emission from V0332+53 in quiescence may indicate that cooling of the NS crust powers its post-outburst luminosity.  This argument is supported by analogy with low-mass X-ray binary NSs in quiescence, where purely thermal emission has been found to be associated with a lack of variability \citep[e.g.][]{Walsh15,Bahramian15}. 

Accretion episodes deposit heat, through density-driven pycnonuclear reactions, in the inner crust (deep crustal heating), which diffuses through the core and slowly leaks to the surface over $\sim10^4-10^5$ years \citep{Haensel90,Brown98}.
Observations of the quiescent luminosity well after an outburst, combined with an estimate of the mass transfer rate, can measure the fraction of heat deposited in the core during outbursts that emerges as X-rays, vs. that escaping as undetected neutrinos \citep[e.g.][]{Colpi01,Yakovlev04,Heinke09}.
We plot the X-ray lightcurve observed for V0332+53 by \textit{RXTE/ASM} between 1995 and 2004 and observed by \textit{Swift} between 2005 and 2016 (Figure.~\ref{fig:lc}). We estimate the time-averaged mass transfer rate from the companion to the neutron star based on the total X-ray luminosity observed, 
\begin{equation}
L_{x} \sim \frac{GM_{NS}\dot{M}}{R_{NS}}
\end{equation}

where we assume $M_{NS}$=1.4 \Msun, and $R_{NS}$=11.5 km. Using the Crab flux and countrate as a reference, we calculate the X-ray flux based on the observed countrate from each instrument.  We compute bolometric corrections assuming a power-law spectrum between 0.1 keV - 100 keV with photon index of 2.  We calculate that the time-averaged mass transfer is $9.8 \times 10^{-11}$ \Msun yr$^{-1}$,  with an uncertainty of $(\pm0.1)\times10^{-11}$ \Msun yr$^{-1}$ from the conversion from countrate to flux, and an uncertainty of $(\pm 0.4) \times 10^{-11}$ \Msun yr$^{-1}$ from our uncertainty in the $M/R$ value.

The thermal radiation that we see in quiescence could have one or more of three origins.  It could be driven entirely by continued accretion, though this seems unlikely (see above).  It could originate from heat deposited in the outer crust during the previous (2004) accretion episode (as suggested by Wijnands \& Degenaar 2016).  Or it could originate from heat deposited in the inner crust over multiple accretion episodes. The thermal conductivity of NSs is altered by $B$ fields reaching $10^{12}$ G, permitting faster transport along magnetic field lines than across them; this predicts that NSs with relatively strong $B$ fields will show hot spots at their magnetic poles \citep{Greenstein83,Potekhin01,Geppert04}.  

Since our observation was taken 4 years after the last major outburst (see, e.g., crust cooling curves in \citealt{Shternin07,Brown09}), we suspect that the thermal flux at this point is dominated by heat stored in the core. However, we do not know the $B$ field and temperature distributions over the surface of the NS, and cannot yet rule out that a large portion of the emitted heat is radiated uniformly from the NS surface. Therefore, we calculate two reasonable limits on the thermal luminosity from deep crustal heating; one from the {\texttt nsa[pole]} hot spot model, and the other from the {\texttt nsa[pole]+nsa} hot spot plus surface model.  

\begin{figure}
	\includegraphics[width=0.45\textwidth]{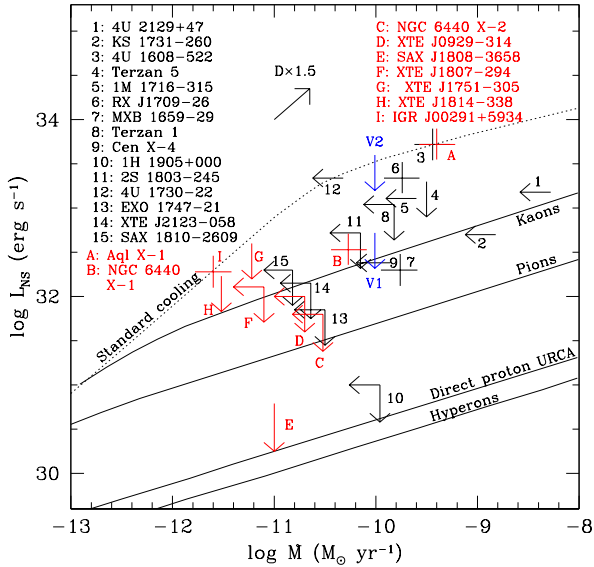}
    \caption{Time-averaged mass transfer rate (in \Msun/yr) vs. quiescent thermal NS luminosity, for NS low-mass X-ray binaries (red: accreting millisecond pulsars; black: other NSs; after \citealt{Heinke07,Heinke09}) and (in blue) for V0332+53. V1 marks the upper limit on thermal luminosity for the hot spot plus surface spectral model, while V2 marks the upper limit for the hot spot model alone. Sample calculations of standard and enhanced neutrino cooling are plotted following \citet{Yakovlev04}; see also \citet{Wijnands13}. }
    \label{fig:spec}
\end{figure}

Comparing the time-averaged mass transfer rate with the observed quiescent thermal NS luminosity allows identification of a NS as following ``standard'' cooling tracks (dominated by modified Urca and/or neutron-neutron neutrino bremssstrahlung; \citealt{Yakovlev04,Page06}) or ``enhanced'' cooling tracks (e.g. direct Urca processes involving protons or hyperons, or similar processes involving pions, kaons, or quark matter; \citealt{Yakovlev04}).  Higher-mass NSs will have higher central densities, possibly permitting ``enhanced'' cooling, while lower-mass NSs are likely to follow ``standard'' cooling \citep{Beznogov15}.  NSs in HMXBs have not had time to accrete substantial mass, but can be born with a range of masses \citep[e.g.][]{Rawls11}, so there is not a theoretical preference for the cooling rate of V0332+53.

We see (Figure ~\ref{fig:spec}) that if V0332+53's thermal luminosity is dominated by the hotspots that we have detected, then the NS is relatively cool, and seems to require slightly enhanced cooling. Alternatively, the rate of deep crustal heating per accretion episode might be reduced for young NSs like this one, since the relevant layers of the crust may not have been completely replaced by accreted material; see \citealt{Wijnands13}.   However, if V0332+53's thermal luminosity is dominated by emission from its entire surface, then it can be consistent with standard cooling tracks. The $N_H$ to V03332+53 (as for most HMXBs) is unfortunately high enough to prevent tight constraints on the thermal luminosity from the full surface of the NS.

\section{Conclusions}

We performed a 39 ks \textit{XMM-Newton} observation of the transient HMXB V0332+53 in quiescence to investigate the source of the quiescent X-ray emission. The low observed X-ray luminosity, $L_{\textsl{x}}$(0.5-10 keV)$ \sim 4\times 10^{32}$ erg s $^{-1}$, together with its 4.35 second spin period, and $3\times10^{12}$ G magnetic field, indicates that V0332+53 is in the propeller regime.
We attempted a search for pulsations, but the small number of photons from V0332+53 prevented a useful upper limit.

 The X-ray spectrum of V0332+53 is quite soft, indicating thermal radiation from the surface, but the inferred radius when fit with a hydrogen NS atmosphere model is much smaller than the NS radius (a simple analysis suggests 0.27$^{+0.22}_{-0.10}$ km), requiring hot spots on the NS. 

From the known outburst history of V0332+53, we estimated the time-averaged mass transfer rate as $9.8\pm0.4 \times 10^{-11}$ \Msun yr$^{-1}$.  If the hotspot emission from V0332+53 is the dominant thermal emission from this NS, then this suggests that V0332+53 exhibits enhanced neutrino cooling, or reduced levels deep crustal heating due to the crust still being partly primordial, rather than accreted \citep{Wijnands13}. However, we can only place an upper limit of $L_{bol}< 3 \times 10^{33}$ on thermal emission from the entire surface of the NS; if there is a strong thermal emission component from the full surface (in addition to the hotspot emission), then V0332+53 is consistent with standard cooling rates.

\section*{Acknowledgements}

COH acknowledges support from an NSERC Discovery Grant, and discussions with Silas Laycock, Rudy Wijnands, and Nathalie Degenaar.

\bibliographystyle{mnras}
\bibliography{Bibliography}

\bsp	% typesetting comment
\label{lastpage}
\end{document}